\documentclass[letterpaper]{article} 
\usepackage{aaai2027}  
\usepackage[hyphens]{url}  
\usepackage{graphicx} 
\usepackage{natbib}  
\usepackage{caption} 
\usepackage{algorithm}
\usepackage{algorithmic}

\usepackage{newfloat}
\usepackage{listings}
\DeclareCaptionStyle{ruled}{labelfont=normalfont,labelsep=colon,strut=off} 
\floatstyle{ruled}
\newfloat{listing}{tb}{lst}{}
\floatname{listing}{Listing}

\usepackage{booktabs}

\usepackage{amsmath,amssymb}
\usepackage{multirow}
\usepackage{makecell}
\usepackage{pifont}

\newcommand{\gainpos}[1]{\textnormal{(+#1)}}
\newcommand{\gainneg}[1]{\textnormal{(-#1)}}
\newcommand{\gainzero}{\textnormal{(+0.00)}}

\newcommand{\sgainpos}[2]{%
  \makecell[c]{#1\\[0.15ex]\gainpos{#2}}}

\newcommand{\sgainneg}[2]{%
  \makecell[c]{#1\\[0.15ex]\gainneg{#2}}}

\newcommand{\sgainzero}[1]{%
  \makecell[c]{#1\\[0.15ex]\gainzero}}

\newcommand{\bestgainpos}[2]{%
  \makecell[c]{\textbf{#1}\\[0.15ex]\gainpos{#2}}}

\newcommand{\secondgainpos}[2]{%
  \makecell[c]{\underline{#1}\\[0.15ex]\gainpos{#2}}}

\newcommand{\secondgainneg}[2]{%
  \makecell[c]{\underline{#1}\\[0.15ex]\gainneg{#2}}}

\newcommand{\hgainpos}[2]{\textbf{#1}\,(+#2)}

\title{
\makebox[\textwidth][c]{%
\raisebox{-0.30\height}{%
\includegraphics[
    width=1.35cm,
    height=1.35cm,
    keepaspectratio
]{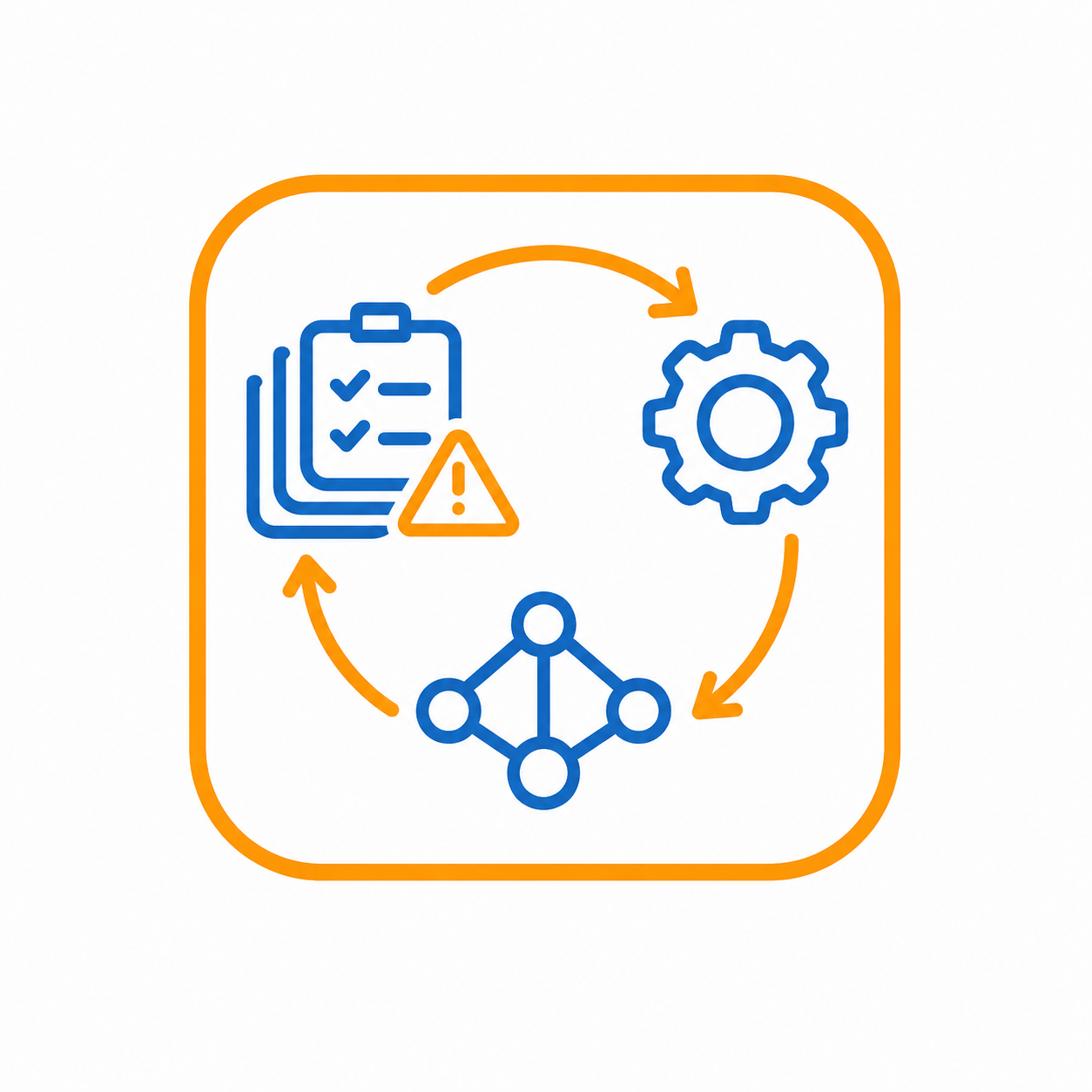}}%
\hspace{0.8em}%
\parbox[c]{0.57\textwidth}{%
\centering
\textit{Living-Harness} Is an Interactive-Agent Evolver
}%
}}
\author{
    Yuetian Du\textsuperscript{\rm 1}\equalcontrib,
    Yucheng Wang\textsuperscript{\rm 1}\equalcontrib,
    He Xu\textsuperscript{\rm 1},
    Jiexu Xu\textsuperscript{\rm 4},
    Shanwen Tan\textsuperscript{\rm 1},
    Bing Zhao\textsuperscript{\rm 2},\\
    Boyu Yang\textsuperscript{\rm 2},
    Zhijie Xu\textsuperscript{\rm 5},
    Ming Kong\textsuperscript{\rm 1},
    Hu Wei\textsuperscript{\rm 2}\corresponding,
    Jie Liu\textsuperscript{\rm 3}\corresponding,
    Qiang Zhu\textsuperscript{\rm 1}\corresponding
}
\affiliations{
    \textsuperscript{\rm 1}Zhejiang University,
    \textsuperscript{\rm 2}Alibaba Group,
    \textsuperscript{\rm 3}City University of Hong Kong\\
    \textsuperscript{\rm 4}The Hong Kong University of Science and Technology,
    \textsuperscript{\rm 5}University of Michigan\\
    \{duyuetian2002,zhuq\}@zju.edu.cn,
    jliu.ee@my.cityu.edu.hk,
    jxu130@connect.hkust-gz.edu.cn
}

\begin{document}

\nocopyright

\maketitle

\begin{abstract}
Large language model (LLM) agents may recover from a failure within an episode or after a retry, yet the same execution failure can recur in later tasks because post-episode feedback rarely revises the persistent harness that guides future interactions. Static harnesses improve reliability through fixed tools, context, memory, and workflow structures, but remain unchanged after deployment. We propose \textbf{Living-Harness}, a self-evolving agent harness that converts each completed trajectory and its evaluator signals into posterior evidence for bounded harness updates. Guided by a domain-level \textbf{Evolution-SOP} (\textbf{S}tandard \textbf{O}perating \textbf{P}rocedure), Living-Harness extracts an episode abstraction and structured update evidence, and writes two complementary forms of procedural knowledge: episodic memory that records trigger conditions, failure patterns, and recovery actions, and a state graph that records state nodes, repair edges, and transition rules. The updated harness state is retrieved to guide future interactions, while tools and base context remain frozen, allowing procedural repairs to accumulate across evolution cycles. On eight interactive environments derived from $\tau^2$-Bench and MultiWOZ-2.4, Living-Harness improves average Pass@1 over the strongest interactive baseline by 10.07 and 9.91 percentage points, respectively, and supports retrieval-only reuse of the evolved harness state across model backbones. Our code will be made publicly available soon at \url{https://github.com/anotherbricki/Living-Harness}.

\end{abstract}


\section{Introduction}

\begin{figure}
    \centering
    \includegraphics[width=\linewidth]{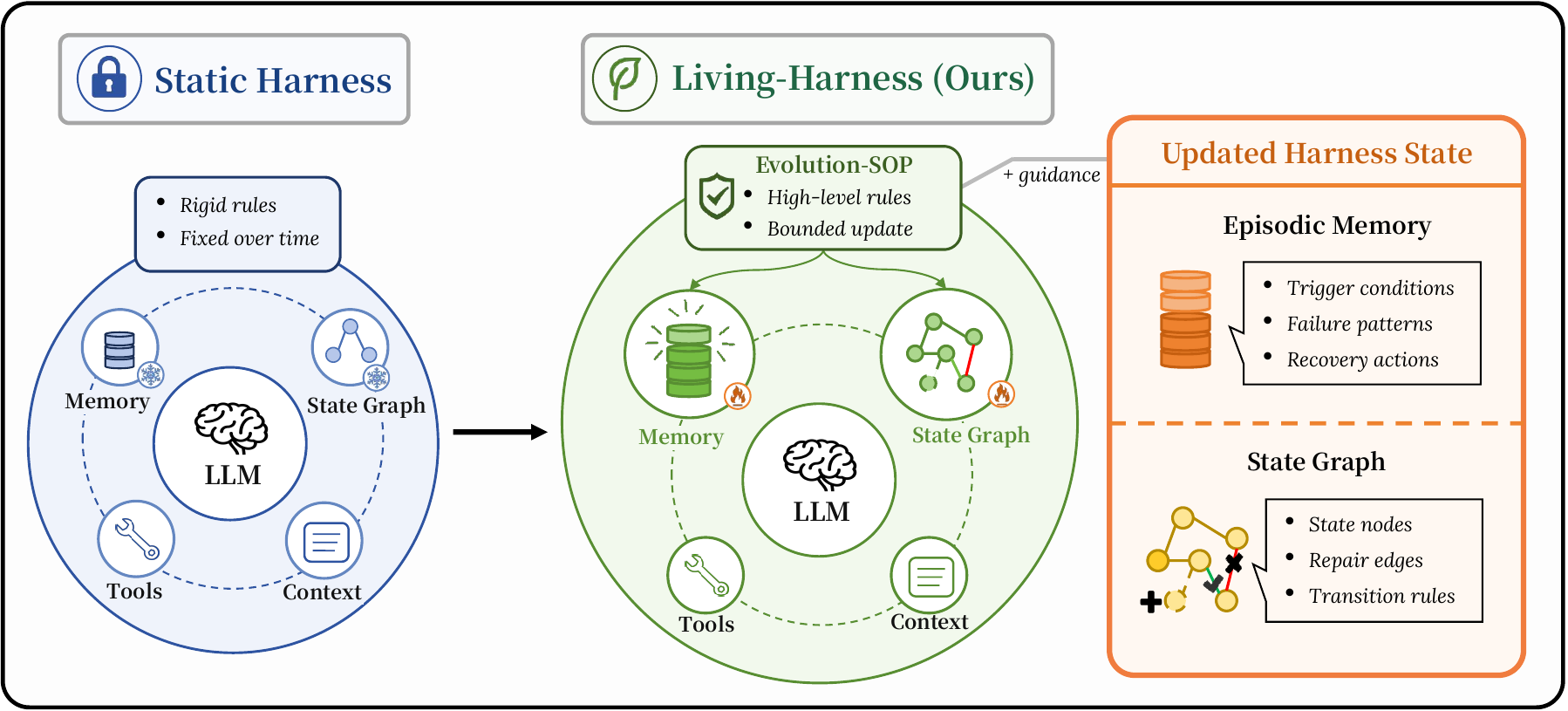}
    
    \caption{Static vs. living harnesses. A static harness remains fixed after deployment. Living-Harness uses an Evolution-SOP to guide bounded updates to episodic memory and the state graph for future retrieval, while tools and base context remain frozen.}
    \label{fig:fig1}
\end{figure}

Large language model (LLM) agents are increasingly deployed as decision-making interfaces for task-oriented dialogue and tool-mediated environments~\cite{DBLP:journals/fcsc/WangMFZYZCTCLZWW24,DBLP:journals/chinaf/XiCGHDHZWJZZFWXZWJZLYDW25,budzianowski-etal-2018-multiwoz,DBLP:journals/corr/abs-2406-12045}. Although an agent may recognize an error from feedback or recover after a task-local retry, the same execution failure can recur in later interactions because the correction often disappears with the completed episode. For example, an agent may correctly conclude that a user should be transferred to a human operator, yet still fail to invoke the required transfer tool when a similar situation appears again. This exposes a gap between correcting an individual response and repairing the persistent procedure that guides future behavior. Practical agent systems address procedural reliability through an external \emph{harness}, which organizes prompts, tools, context, memory, workflows, and evaluation interfaces around the base model~\cite{meng2026agentharness,DBLP:journals/corr/abs-2603-28052}. However, as illustrated in Figure~\ref{fig:fig1}, a static harness is fixed after deployment: it can impose predefined rules, but it cannot incorporate newly observed failure patterns and recovery actions into its future procedure. This leads to a concrete question: \textit{how can an agent turn post-episode failures into persistent procedural repairs?}


Existing methods address parts of this problem from two directions. Harness-design and workflow-optimization methods construct stronger prompts, SOPs, tool constraints, skills, or execution structures that keep an agent within predefined paths~\cite{meng2026agentharness,DBLP:journals/corr/abs-2603-28052,DBLP:journals/corr/abs-2501-09316,DBLP:conf/emnlp/XiongSDZSWL25,DBLP:conf/iclr/ZhangXYTCCZCHWZ25}. They demonstrate the value of external procedural structure, but the resulting harness is commonly designed or optimized before deployment and then reused as a fixed artifact. Reflection and experience-memory methods instead retain critiques, summaries, or successful and failed trajectories for subsequent attempts~\cite{DBLP:conf/nips/ShinnCGNY23,DBLP:conf/aaai/Zhao0XLLH24,DBLP:journals/corr/abs-2509-25140}. These methods provide useful experience, but a textual lesson such as ``transfer the user to a human'' does not by itself specify the trigger condition, required tool action, and workflow transition needed to prevent the same failure in future episodes. The missing capability is therefore not feedback collection alone, but \emph{persistent procedural repair}: converting evaluator-grounded failure evidence into a reusable relation between when a failure occurs, what action or transition is missing, and how a later rollout should recover.


A useful procedural-repair mechanism must satisfy several requirements. First, an update should be grounded in a completed trajectory and its evaluation signals, rather than in an unverified self-critique alone. Second, the repair must persist across episodes while remaining scoped to the relevant task family or failure condition. Third, adaptation should modify only the stateful parts of the harness, leaving the available tools and base context frozen so that accumulated experience does not freely rewrite the agent's operational boundaries. Finally, the persistent state should capture two complementary forms of knowledge: \emph{experiential knowledge} about why a failure occurred and how it was recovered, and \emph{workflow knowledge} about which state-conditioned action or transition should be introduced or revised. These requirements motivate an evolving harness state composed of complementary episodic memory and state-graph structures.


We therefore propose \textbf{Living-Harness}, a self-evolving harness that converts evaluated interaction trajectories into persistent procedural repairs. As shown in Figure~\ref{fig:fig1}, the tools and base context remain frozen, while episodic memory and the state graph evolve. After each episode, an evaluator provides feedback on the completed interaction, and a domain-level \textbf{Evolution-SOP} guides how failures are interpreted and which repairs are committed to the harness state. Episodic memory records trigger conditions, failure patterns, and recovery actions, while the state graph records state nodes, repair edges, and transition rules. In future episodes, relevant memory and graph entries are retrieved as procedural context, allowing recurring failures to progressively improve subsequent interactions.

To formalize this cross-episode adaptation, we view the agent as reasoning under uncertainty not only about the environment, but also about the procedural knowledge available to future decisions. This program-state POMDP perspective distinguishes within-episode execution from post-episode harness revision and motivates our rollout--evaluate--update formulation.


The main contributions of this paper are summarized as follows:
\begin{itemize}
    \item \textbf{Persistent procedural repair for interactive agents.}
    We identify a gap between task-local correction and cross-episode adaptation: feedback may improve an individual retry without installing a reusable repair into the harness that guides later interactions. We formulate the target of adaptation as evaluator-grounded procedural repairs that persist beyond the episode in which a failure is observed.

    \item \textbf{The Living-Harness framework.}
    We introduce a rollout--evaluate--update framework in which a domain-level Evolution-SOP converts completed trajectories and evaluation signals into bounded updates of episodic memory and a state graph. These components store complementary experiential and workflow knowledge, are retrieved to guide future episodes, and evolve while the tools and base context remain frozen. We further provide a program-state POMDP interpretation of this cross-episode update process.

    \item \textbf{Evaluation of recovery accumulation and harness reuse.}
    We evaluate Living-Harness on eight interactive environments derived from \(\tau^2\)-Bench and MultiWOZ-2.4. It improves average Pass@1 over the strongest interactive baseline by 10.07 and 9.91 percentage points, respectively, while exhibiting cycle-wise recovery accumulation and retrieval-only reuse of the evolved harness state across model backbones.
\end{itemize}




\section{Related Work}

\noindent\textbf{Interactive LLM Agents.} Research on interactive LLM agents examines how language models act over multiple turns by conditioning on observations, user feedback, tool outputs, and task constraints rather than producing isolated one-shot responses. Task-oriented dialogue benchmarks provide an early formulation of this setting, requiring agents to maintain dialogue state, follow domain policies, and complete user goals, as formalized by MultiWOZ \cite{budzianowski-etal-2018-multiwoz,DBLP:conf/sigdial/0001MY22} and Schema-Guided Dialogue \cite{DBLP:conf/aaai/RastogiZSGK20}; recent tool-agent-user benchmarks extend it to realistic domains requiring both tool use and policy compliance \cite{DBLP:journals/corr/abs-2406-12045}. LLM agents broaden this line to reasoning, tool use, coordination, and trial-level adaptation, as summarized in recent surveys on LLM-based autonomous agents \cite{DBLP:journals/fcsc/WangMFZYZCTCLZWW24,DBLP:journals/chinaf/XiCGHDHZWJZZFWXZWJZLYDW25}. Representative systems include WebGPT \cite{DBLP:journals/corr/abs-2112-09332} for browsing-grounded answering, ReAct \cite{DBLP:conf/iclr/YaoZYDSN023} for reasoning-action interleaving, Toolformer \cite{DBLP:conf/nips/SchickDDRLHZCS23} and ToolLLM \cite{DBLP:journals/csur/QinHLCDCZZHXHFSWQTZLSXZ25} for API use, AutoGen \cite{DBLP:journals/corr/abs-2308-08155} for multi-agent coordination, and Reflexion \cite{DBLP:conf/nips/ShinnCGNY23} for feedback-based retry. These methods improve behavior within an interaction or retry, whereas Living-Harness turns post-episode evidence into persistent procedural repairs for future episodes.

\begin{figure*}[t]
    \centering
    \includegraphics[width=\linewidth]{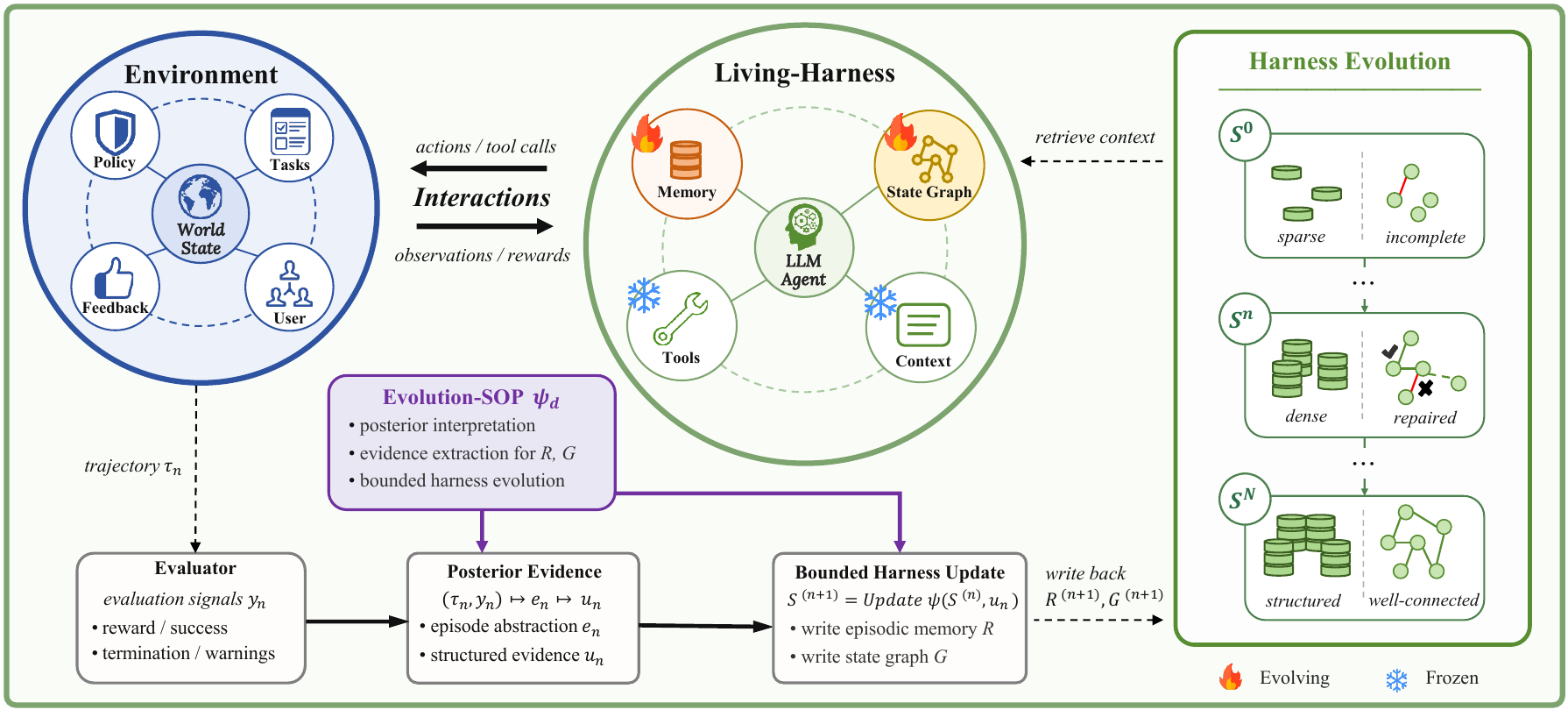}
    \caption{
    Overview of Living-Harness. In episode \(n\), the agent interacts with the environment under harness state \(S^{(n)}\), producing trajectory \(\tau_n\) and evaluation signals \(y_n\). Guided by the Evolution-SOP \(\psi_d\), these signals are converted into posterior evidence and used to update episodic memory \(\mathcal{R}\) and the state graph \(G\). The resulting state \(S^{(n+1)}\) is written back and retrieved to guide future episodes, while tools and base context remain frozen.
    }
    \label{fig:fig2}
\end{figure*}

\noindent\textbf{Agent Harnesses.} As LLM agents move from isolated prompting to tool use, memory, evaluation, and multi-step execution, their behavior is increasingly shaped by the external program surrounding the model. We use \emph{agent harness} to denote this control layer, which externalizes prompts, tools, memory, workflows, execution loops, constraints, tracing, and evaluation interfaces \cite{meng2026agentharness,DBLP:journals/corr/abs-2603-28052}. Studies of agent evaluation \cite{DBLP:journals/tmlr/KapoorSSNN25}, context engineering \cite{DBLP:journals/corr/abs-2601-21557}, meta-tooling \cite{DBLP:journals/corr/abs-2601-22037}, workflow optimization \cite{DBLP:conf/iclr/HuLC25,DBLP:conf/iclr/ZhangXYTCCZCHWZ25}, guardrails, sessions, tracing, and skill \cite{DBLP:journals/corr/abs-2602-12670,DBLP:journals/corr/abs-2601-04748,DBLP:journals/corr/abs-2602-16653,DBLP:journals/corr/abs-2604-02268} benchmarks further show that this surrounding structure strongly affects practical reliability. Skills are a central harness-level abstraction because they package procedural knowledge into reusable, invocable units, including instructions, code, resources, applicability conditions, and interfaces that can be loaded or selected at inference time \cite{DBLP:journals/corr/abs-2602-12430,DBLP:journals/corr/abs-2602-20867,DBLP:journals/corr/abs-2602-08004,DBLP:journals/tmlr/WangX0MXZFA24}. Existing harness and skill systems provide reusable procedural structure, but are commonly specified or optimized before deployment. Living-Harness instead updates its episodic memory and state graph from evaluated post-episode evidence while keeping tools and base context fixed.

\noindent\textbf{Self-Evolving Agent Systems.} Self-evolving agent systems convert interaction experience into persistent artifacts such as memory, prompts, skills, workflows, policies, or architectures \cite{DBLP:journals/tmlr/GaoGHHJLLQQRWWXZZZXFZLQ26,DBLP:journals/corr/abs-2508-07407,DBLP:journals/corr/abs-2604-17091}. Reflection \cite{DBLP:conf/nips/ShinnCGNY23} and reasoning-memory methods \cite{DBLP:conf/aaai/Zhao0XLLH24,DBLP:journals/corr/abs-2509-25140} store feedback or trajectories as reusable lessons or retrievable memories; skill-centered systems acquire, route, or evolve reusable capabilities \cite{DBLP:journals/corr/abs-2603-02766}, from open-ended skill libraries \cite{DBLP:journals/tmlr/WangX0MXZFA24} to hierarchical skill banks \cite{DBLP:journals/corr/abs-2602-08234}, evolving memory skills \cite{DBLP:journals/corr/abs-2602-02474}, and skill-transfer routers \cite{DBLP:journals/corr/abs-2602-19672}; and workflow, architecture, and environment-evolution methods optimize higher-level execution structures such as code-represented workflows \cite{DBLP:conf/iclr/ZhangXYTCCZCHWZ25}, modular tool-use policies \cite{DBLP:conf/acl/YangHMLMH26}, multi-agent architectures \cite{DBLP:conf/icml/ZhangNF00025}, synthetic environments \cite{DBLP:journals/corr/abs-2604-18292}, and test-time procedural strategies \cite{DBLP:journals/corr/abs-2604-15097}. These approaches typically evolve a particular artifact, such as a memory, skill, policy, workflow, or architecture. Living-Harness instead converts evaluated failures into coordinated updates of episodic memory and workflow structure within a persistent harness state.



\section{Method}
\label{sec:method}


This section presents \textbf{Living-Harness}, a self-evolving agent harness that accumulates persistent procedural repairs through a rollout--evaluate--update loop. As illustrated in Figure~\ref{fig:fig2}, the agent retrieves relevant entries from the current harness state and interacts with the environment under fixed tools and base context. After the episode is evaluated, the \textbf{Evolution-SOP} converts the completed interaction and its evaluation signals into structured evidence for updating episodic memory and the state graph. The updated state is written back for retrieval in later episodes, allowing procedural repairs to accumulate across interactions.

\subsection{Problem Setting and Harness State}

We consider an LLM agent operating in a domain \(d\) over a sequence
of interaction episodes. In episode \(n\), the agent receives a task
\(x_n\), interacts with the environment, and produces a trajectory
\(\tau_n\) consisting of observations, responses, and tool actions.
After the interaction terminates, an evaluator produces signals
\(y_n\) describing its execution outcome. The goal is to use
\((\tau_n,y_n)\) to improve the procedural guidance available to
later episodes.

Living-Harness separates fixed execution resources from an evolving
procedural state. At episode \(n\), the harness is represented as
\begin{equation}
H_d^{(n)}
=
\left(C_d^{\mathrm{act}},\psi_d,S_d^{(n)}\right),
\qquad
S_d^{(n)}
=
\left(\mathcal{R}_d^{(n)},G_d^{(n)}\right).
\end{equation}
Here, \(C_d^{\mathrm{act}}\) contains the tools, base context, and
domain rules available to the actor; \(\psi_d\) is the fixed
Evolution-SOP governing post-episode updates; and \(S_d^{(n)}\) is
the evolving harness state. The state remains fixed throughout the
current episode and may be revised only after the completed
interaction has been evaluated. Living-Harness therefore updates
neither the base model nor \(C_d^{\mathrm{act}}\); its adaptation
target is \(S_d^{(n)}\).

The evolving state contains two complementary forms of procedural
knowledge. Episodic memory \(\mathcal{R}_d^{(n)}\) stores scoped
records of trigger conditions, failure patterns, and recovery actions.
The state graph \(G_d^{(n)}\) stores state nodes, transition rules,
and repair edges that connect observed conditions to missing or
revised actions. During later episodes, relevant entries from both
components are retrieved and rendered as procedural context, while
the fixed rules in \(C_d^{\mathrm{act}}\) retain precedence.

Conceptually, \(S_d^{(n)}\) induces an episode-level program state
\(z^{(n)}\): the environment evolves within an episode, whereas the
procedural information in \(z^{(n)}\) changes only through
post-episode harness updates. This two-timescale view is formalized
in the Program-State Formulation.

\noindent\textbf{Initialization.}
The initial state contains no episodic memory and only a coarse graph
scaffold:
\begin{equation}
S_d^{(0)}
=
\left(\emptyset,G_{d,\mathrm{scaf}}\right).
\end{equation}
The scaffold provides domain roots and task-family structure but
contains no fine-grained failure repairs. Such repairs are induced
from evaluated interaction experience through the Evolution-SOP.

\begin{table*}[t]
\centering
\setlength{\tabcolsep}{3.5pt}
\begin{tabular*}{0.95\textwidth}{
  @{\extracolsep{\fill}}lcccccccc@{}
}
\toprule
& \multicolumn{4}{c}{$\tau^2$-Bench}
& \multicolumn{4}{c}{MultiWOZ-2.4} \\
\cmidrule(lr){2-5}
\cmidrule(lr){6-9}
Model
& Retail
& Airline
& Telecom
& Average
& \makecell[c]{1-Domain}
& \makecell[c]{2-Domains}
& \makecell[c]{3-Domains}
& Average \\
\midrule

\multicolumn{9}{l}{\textit{Flagship Models}} \\
\midrule

Gemini 3 Pro
& 75.88
& 80.50
& \textbf{91.01}
& \underline{82.92}
& \textbf{79.20}
& 54.52
& \underline{24.48}
& \underline{55.80} \\
GLM-5
& 73.68
& \underline{82.50}
& \underline{86.84}
& 80.66
& 46.02
& 21.24
& 0.00
& 23.80 \\
Qwen3-max
& 72.20
& 59.50
& 84.20
& 74.85
& 37.17
& 35.97
& 0.00
& 31.10 \\
GPT-5.2
& 57.02
& 70.00
& 52.23
& 57.39
& 48.23
& 37.88
& 7.69
& 35.90 \\
Kimi-k2
& 70.60
& 56.50
& 65.80
& 66.11
& 54.88
& 41.56
& 0.00
& 38.63 \\
\midrule

\multicolumn{9}{l}{\textit{Interactive Baselines (GPT-5.2)}} \\
\midrule

ReAct
& 52.63
& 66.00
& 53.86
& 55.54
& 70.35
& 19.18
& 9.09
& 29.30 \\
Reflexion
& \underline{78.07}
& 80.00
& 64.91
& 73.02
& \underline{77.88}
& 51.35
& 21.68
& 53.10 \\
AWM
& 56.14
& 66.00
& 39.47
& 51.08
& 67.26
& 40.57
& 13.99
& 42.80 \\
ReasoningBank
& 59.65
& 70.00
& 37.72
& 52.52
& 77.27
& \underline{58.33}
& 7.50
& 55.59 \\
EvoTest
& 57.31
& 64.00
& 38.07
& 50.62
& 62.39
& 34.86
& 11.89
& 37.80 \\
\midrule

\multicolumn{9}{l}{\textit{Ours (GPT-5.2)}} \\
\midrule

\textbf{Living-Harness}
& \textbf{85.96}
& \textbf{88.00}
& 78.07
& \textbf{83.09}
& 76.55
& \textbf{70.52}
& \textbf{25.87}
& \textbf{65.50} \\
\bottomrule
\end{tabular*}

\caption{Main results on $\tau^2$-Bench and MultiWOZ-2.4.
All values are Pass@1 (\%).
Averages are weighted by the number of evaluated tasks.
Bold and underlined values indicate the best and second-best results,
respectively.}
\label{tab:main_results}
\end{table*}

\subsection{Evolution-SOP for Procedural Repair}

The \textbf{Evolution-SOP} \(\psi_d\) is a fixed, domain-level
protocol that governs how evaluated interactions revise the evolving
harness state. It operates after rollout rather than serving as part
of the actor's task-execution prompt. Across domains, Living-Harness
uses the same posterior--extract--commit procedure, while
\(\psi_d\) provides domain-specific guidance for interpreting
failures, assigning update scope, and respecting domain and tool
constraints.

Given a task \(x_n\), its completed trajectory \(\tau_n\), evaluator
signals \(y_n\), and the fixed execution resources
\(C_d^{\mathrm{act}}\), the Evolution-SOP first produces an episode
abstraction:
\begin{equation}
e_n
=
\operatorname{Post}_{M,\psi_d}
\left(
x_n,\tau_n,y_n,C_d^{\mathrm{act}}
\right).
\end{equation}
The abstraction identifies the task objective, verified interaction
facts, execution outcome, and the critical failure or recovery point.
Rather than retaining the full trajectory, it isolates the procedural
information that may remain useful after the current episode ends.

The episode abstraction is then converted into evidence for the two
components of the harness state:
\begin{equation}
u_n
=
\left(
u_n^{\mathcal{R}},
u_n^G
\right)
=
\operatorname{Extract}_{\psi_d}(e_n).
\end{equation}
Here, \(u_n^{\mathcal{R}}\) captures trigger conditions, failure
patterns, and recovery actions for episodic memory, while \(u_n^G\)
captures the corresponding states, actions, and transitions for the
state graph. Thus, the memory preserves why and under what condition
a repair is useful, whereas the graph records where that repair
changes the future procedure.

The extracted evidence is treated as a candidate repair. Before being
committed, it is checked for evidential support, task scope, and
consistency with the fixed domain and tool constraints. Accepted
candidates create or strengthen memory entries and state-conditioned
repair edges; unsupported or conflicting candidates leave the
persistent state unchanged.

\subsection{Rollout--Evaluate--Update Loop}

At episode \(n\), Living-Harness constructs a task-conditioned query
from the current task \(x_n\) and its scope \(f_n\):
\begin{equation}
q_n = Q(x_n,f_n).
\end{equation}
Relevant entries are retrieved from the current episodic memory and
state graph and rendered as actor-facing procedural context:
\begin{equation}
\kappa_n
=
\operatorname{Render}\!\left(
\operatorname{Retrieve}_{\mathcal{R}}
    (\mathcal{R}^{(n)},q_n),
\operatorname{Retrieve}_{G}
    (G^{(n)},q_n)
\right).
\end{equation}
This selective projection exposes the procedural knowledge most
relevant to the current task rather than replaying the complete
interaction history. The agent then interacts with the environment
using the fixed execution resources and the retrieved context:
\begin{equation}
\tau_n
\sim
p_M\!\left(
\tau\mid x_n,C_d^{\mathrm{act}},\kappa_n
\right).
\end{equation}
The harness state remains fixed throughout the rollout.

After the interaction is completed, the evaluator produces
\begin{equation}
y_n=E(x_n,\tau_n).
\end{equation}
The Evolution-SOP converts \(\tau_n\) and \(y_n\) into the episode
abstraction \(e_n\) and candidate repair evidence \(u_n\) defined
above. The accepted evidence is then written into the persistent
harness state:
\begin{equation}
S^{(n+1)}
=
\operatorname{Update}_{\psi_d}
\left(
S^{(n)},u_n;C_d^{\mathrm{act}}
\right).
\end{equation}
The memory update preserves reusable failure--recovery experience,
while the graph update revises the corresponding procedural
transitions. If no candidate repair is accepted, the state remains
unchanged. Otherwise, the updated state becomes available only to
subsequent interactions, completing the rollout--evaluate--update
loop without allowing post-episode evidence to alter the rollout that
produced it.

\noindent\textbf{Task-local correction and persistent evolution.}
For tasks that permit retries, a failed attempt may produce a local
reflection used only within the current task instance. Such
task-local feedback is discarded when the instance ends. Persistent
evolution instead commits evaluator-grounded repairs to episodic
memory and the state graph, making them available to later task
instances.

\subsection{Program-State Analysis}

We analyze Living-Harness through a Partially Observable Markov
Decision Process (POMDP) lens. In a standard POMDP, the environment
state is not directly observed and must instead be inferred from the
interaction history. Living-Harness augments this view with an
episode-level program state representing the procedural information
available from the evolving harness.

\noindent\textbf{Definition 1 (Environment-state belief).}
Let \(s_t\in\mathcal{S}\) denote the environment state at
interaction step \(t\), and let
\(h_t=(o_{\leq t},a_{<t})\) denote the interaction history. The
standard POMDP belief over the environment is
\begin{equation}
b_t(s)
=
P\!\left(s_t=s\mid h_t\right).
\end{equation}

\noindent\textbf{Definition 2 (Program-augmented belief).}
Using the program state \(z^{(n)}\) induced by the current harness
state \(S^{(n)}\), we augment the environment state and define the
corresponding joint belief as
\begin{equation}
\widetilde{s}_t^{(n)}=\left(s_t,z^{(n)}\right),
B_t^{(n)}(s,z)
=
P\!\left(s_t=s,z^{(n)}=z\mid h_t\right).
\end{equation}
The environment-only belief is recovered by marginalization,
\(b_t(s)=\sum_z B_t^{(n)}(s,z)\).

The two states evolve at different timescales. During episode \(n\),
\(z^{(n)}\) remains fixed while the joint belief is updated as
\begin{equation}
B_{t+1}^{(n)}(s',z)\propto
\Omega(o_{t+1}\mid s')
\sum_s T(s'\mid s,a_t,z)B_t^{(n)}(s,z).
\end{equation}
After evaluation, updates to episodic memory and the state graph
induce a new program state for later episodes. Thus, \(t\) indexes
within-episode inference, while \(n\) indexes cross-episode harness
evolution.

\noindent\textbf{Proposition 1 (Information refinement).}
For the nested information sets
\(\mathcal{I}_s=\sigma(h_t)\subseteq
\mathcal{I}_{s,z}=\sigma(h_t,z^{(n)})\), consider an
execution-relevant latent variable \(\xi\in L^2\). Define
\(\mathcal{B}(\mathcal I)
=\mathbb{E}[\operatorname{Var}(\xi\mid\mathcal I)]\).
Then
\(\mathcal{B}(\mathcal{I}_{s,z})
=\mathcal{B}(\mathcal{I}_s)-\Gamma_z
\leq\mathcal{B}(\mathcal{I}_s)\), where
\(\Gamma_z
=\mathbb{E}[\operatorname{Var}(
\mathbb{E}[\xi\mid\mathcal{I}_{s,z}]
\mid\mathcal{I}_s)]\geq0\).
The inequality is strict when \(z\) contains nonzero conditional
information about \(\xi\).

Here, \(\mathcal{B}(\mathcal I)\) denotes the Bayes error bound under
information \(\mathcal I\), while \(\mathcal{I}_{s,z}\) augments the
interaction history with the retrieved memory and graph context
represented by \(z^{(n)}\). Proposition~1 therefore shows that
informative procedural context can reduce the optimal prediction
error relative to interaction history alone.

\section{Experiments}

\begin{table*}[t]
\centering
\small
\renewcommand{\arraystretch}{1.08}
\setlength{\tabcolsep}{3pt}
\begin{tabular*}{0.98\textwidth}{
  @{\extracolsep{\fill}}lcccccccc@{}
}
\toprule
\multirow{2}{*}{Cycle}
& \multicolumn{3}{c}{$\tau^2$-Bench}
& \multicolumn{5}{c}{MultiWOZ-2.4} \\
\cmidrule(lr){2-4}
\cmidrule(lr){5-9}
& Retail
& Airline
& Telecom
& Restaurant
& Hotel
& Train
& Attraction
& Taxi \\
\midrule

0
& \sgainzero{57.02}
& \sgainzero{70.00}
& \sgainzero{57.39}
& \sgainzero{30.89}
& \sgainzero{24.11}
& \sgainzero{43.43}
& \sgainzero{40.15}
& \sgainzero{8.21} \\

1
& \sgainpos{74.56}{17.54}
& \sgainpos{76.00}{6.00}
& \sgainpos{67.54}{10.15}
& \secondgainpos{43.47}{12.58}
& \sgainpos{42.64}{18.53}
& \sgainpos{64.65}{21.22}
& \sgainpos{61.36}{21.21}
& \sgainpos{28.72}{20.51} \\

2
& \secondgainpos{79.82}{5.26}
& \bestgainpos{88.00}{12.00}
& \secondgainpos{71.92}{4.38}
& \sgainneg{41.65}{1.82}
& \secondgainpos{46.95}{4.31}
& \secondgainpos{65.86}{1.21}
& \secondgainpos{63.89}{2.53}
& \secondgainpos{36.41}{7.69} \\

3
& \bestgainpos{85.96}{6.14}
& \secondgainneg{86.00}{2.00}
& \bestgainpos{78.07}{6.15}
& \bestgainpos{44.85}{3.20}
& \bestgainpos{48.22}{1.27}
& \bestgainpos{66.26}{0.40}
& \bestgainpos{65.40}{1.51}
& \bestgainpos{36.92}{0.51} \\

\bottomrule
\end{tabular*}

\caption{Per-cycle performance of Living-Harness.
Parenthesized values report absolute changes from the preceding cycle.
Rest.\ and Attr.\ denote Restaurant and Attraction.
Bold and underlined values indicate the best and second-best cycle
performance within each domain, respectively.}
\label{tab:lh_cycle_scaling}
\end{table*}


\subsection{Experimental Setup}

\noindent\textbf{Evaluation Benchmarks.}
We evaluate Living-Harness on two interactive benchmarks:
\begin{itemize}
    \item \textbf{$\tau^2$-Bench}~\cite{DBLP:journals/corr/abs-2506-07982}: a realistic benchmark for conversational agents with multi-turn interaction, policy constraints, and executable tool use. We use three domains: \textit{Retail}, \textit{Airline}, and \textit{Telecom}.
    \item \textbf{MultiWOZ-2.4}~\cite{DBLP:conf/sigdial/0001MY22}: a corrected multi-domain task-oriented dialogue benchmark. We use five primary domains: \textit{Restaurant}, \textit{Hotel}, \textit{Train}, \textit{Attraction}, and \textit{Taxi}.
\end{itemize}
Across both benchmarks, we report task success rate, denoted as Pass@1, as the primary metric.

\noindent\textbf{Baselines.}
We compare Living-Harness against two groups of baselines:
\begin{itemize}
    \item \textit{Flagship Models}: We evaluate strong proprietary and open-weight backbones, including GPT-5.2~\cite{singh2026openaigpt5card}, Gemini 3 Pro~\cite{google2025gemini3}, GLM-5~\cite{DBLP:journals/corr/abs-2602-15763}, Qwen3-max~\cite{qwen2025qwen3max}, and Kimi-k2~\cite{DBLP:journals/corr/abs-2507-20534}. These models are evaluated without online harness evolution to measure the base capability of each agent backbone.
    \item \textit{Interactive Baselines}: We compare against representative interactive and self-improving agent methods, including ReAct~\cite{DBLP:conf/iclr/YaoZYDSN023}, Reflexion~\cite{DBLP:conf/nips/ShinnCGNY23}, Agent Workflow Memory (AWM)~\cite{DBLP:conf/icml/WangMFN25}, ReasoningBank~\cite{DBLP:journals/corr/abs-2509-25140}, and EvoTest~\cite{DBLP:journals/corr/abs-2510-13220}. 
\end{itemize}

\noindent\textbf{Implementation Details.}
All GPT-5.2-based interactive methods, including Living-Harness, use GPT-5.2 with medium reasoning effort as the backbone; within Living-Harness, the actor and all self-evolving modules share this same backbone. For $\tau^2$-Bench, the simulated user is implemented with GPT-5.1. Across the actor and all evolution modules, we use a sampling
temperature of \(0.2\) and set nucleus sampling to
\(\texttt{top\_p}=1.0\). Where applicable, all interactive baselines use the same tool interface, evaluator, and task-level retry budget. We enable task-local reflexion with at most 3 trials per task; this local buffer is used only within the current instance and is never directly written into the global episodic memory \(\mathcal{R}^{(n)}\) or state graph \(G^{(n)}\). Both memory and graph retrieval use top-\(k=3\). 

\subsection{Main Results}
\noindent\textbf{Strong overall gains.} 
Table~\ref{tab:main_results} shows that Living-Harness achieves the best overall performance on both benchmarks, reaching 83.09 average Pass@1 on \(\tau^2\)-Bench and 65.50 on MultiWOZ-2.4. On \(\tau^2\)-Bench, it slightly surpasses the strongest flagship-model average, Gemini 3 Pro at 82.92, despite using GPT-5.2 medium as the base model. It also substantially improves over GPT-5.2-based interactive baselines: compared with Reflexion, the strongest \(\tau^2\)-Bench interactive baseline, Living-Harness improves the average from 73.02 to 83.09. On MultiWOZ-2.4, it improves over the strongest interactive-baseline average, ReasoningBank at 55.59, by nearly 10 points, and outperforms Reflexion by 12.40 points. Although its one-domain score is slightly below Reflexion, Living-Harness achieves the best scores in the two- and three-domain groups, suggesting stronger benefits when procedural repairs must transfer across domain boundaries. These results support our claim that bounded harness-state updates provide gains beyond stronger single-step reasoning or task-local reflection.

\noindent\textbf{Consistent self-evolution.} 
Table~\ref{tab:lh_cycle_scaling} further examines the self-evolution dynamics of Living-Harness. On \(\tau^2\)-Bench, all three domains improve substantially over Cycle 0: Retail rises from 57.02 to 85.96, Telecom from 57.39 to 78.07, and Airline reaches 88.00 at Cycle 2 before remaining high at 86.00 in Cycle 3. MultiWOZ-2.4 shows similarly strong gains, with final-cycle improvements of \(+13.96\) on Restaurant, \(+24.11\) on Hotel, \(+22.83\) on Train, \(+25.25\) on Attraction, and \(+28.71\) on Taxi. The largest gains usually appear after the first evolution cycle, while later cycles produce smaller refinements and occasional mild fluctuations. This pattern is consistent with bounded program evolution: early updates repair missing workflow steps, and later cycles refine the accumulated harness state rather than simply adding more context. 

\begin{figure}[t]
    \centering
    \includegraphics[width=\linewidth]{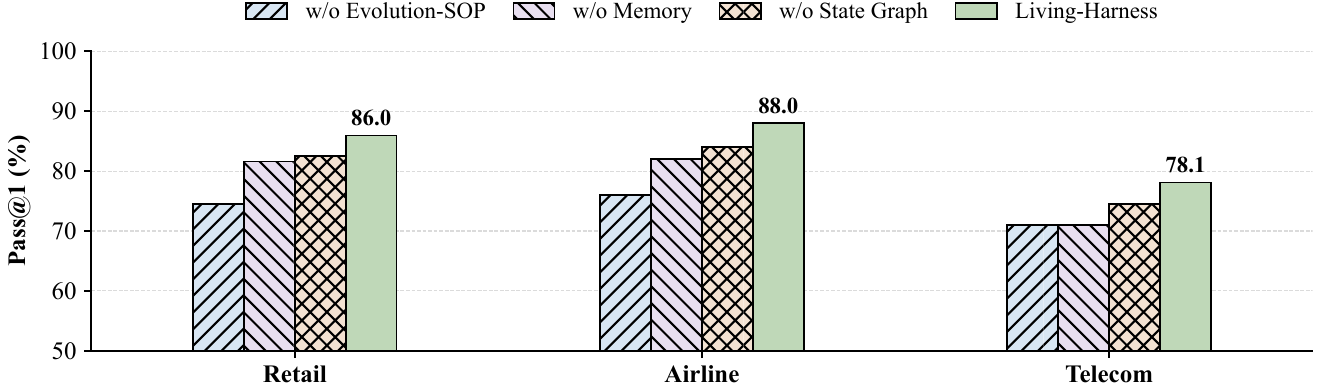}
   
   \caption{Component ablation on $\tau^2$-Bench: removing the Evolution-SOP causes the largest drop, while memory and state graph ablations show complementary contributions.}
    \label{fig:bar_chart}
\end{figure}

\begin{table}[t]
\centering
\small
\setlength{\tabcolsep}{2pt}
\begin{tabular*}{0.99\columnwidth}{
  @{\extracolsep{\fill}}lccccc@{}
}
\toprule
Model
& Rest.
& Hotel
& Train
& Attr.
& Taxi \\
\midrule

Gemini 3 Pro
& 47.83
& 45.18
& 55.35
& 63.64
& 30.26 \\
+ \textbf{Living-Harness}
& \textbf{66.13}
& \textbf{63.71}
& \textbf{66.87}
& \textbf{66.67}
& \textbf{68.72} \\
\midrule

GLM-5
& 24.94
& 16.50
& 18.18
& 27.27
& 0.00 \\
+ \textbf{Living-Harness}
& \textbf{39.36}
& \textbf{38.83}
& \textbf{38.38}
& \textbf{38.13}
& \textbf{43.08} \\
\midrule

Qwen3-max
& 19.45
& 31.98
& 37.98
& 35.10
& 0.00 \\
+ \textbf{Living-Harness}
& \textbf{45.77}
& \textbf{44.42}
& \textbf{47.27}
& \textbf{44.70}
& \textbf{45.13} \\
\midrule

Kimi-k2
& 40.27
& 35.37
& 38.42
& 37.01
& 0.00 \\
+ \textbf{Living-Harness}
& \textbf{44.85}
& \textbf{43.65}
& \textbf{47.27}
& \textbf{47.22}
& \textbf{45.13} \\

\bottomrule
\end{tabular*}

\caption{Retrieval-only cross-model transfer of the frozen
Living-Harness state on MultiWOZ-2.4.
Bold values denote scores obtained with the transferred harness state.
Rest.\ and Attr.\ denote Restaurant and Attraction.}
\label{tab:lh_cross_model_transfer}
\end{table}


\begin{figure*}[t]
    \centering
    \includegraphics[width=0.97\linewidth]{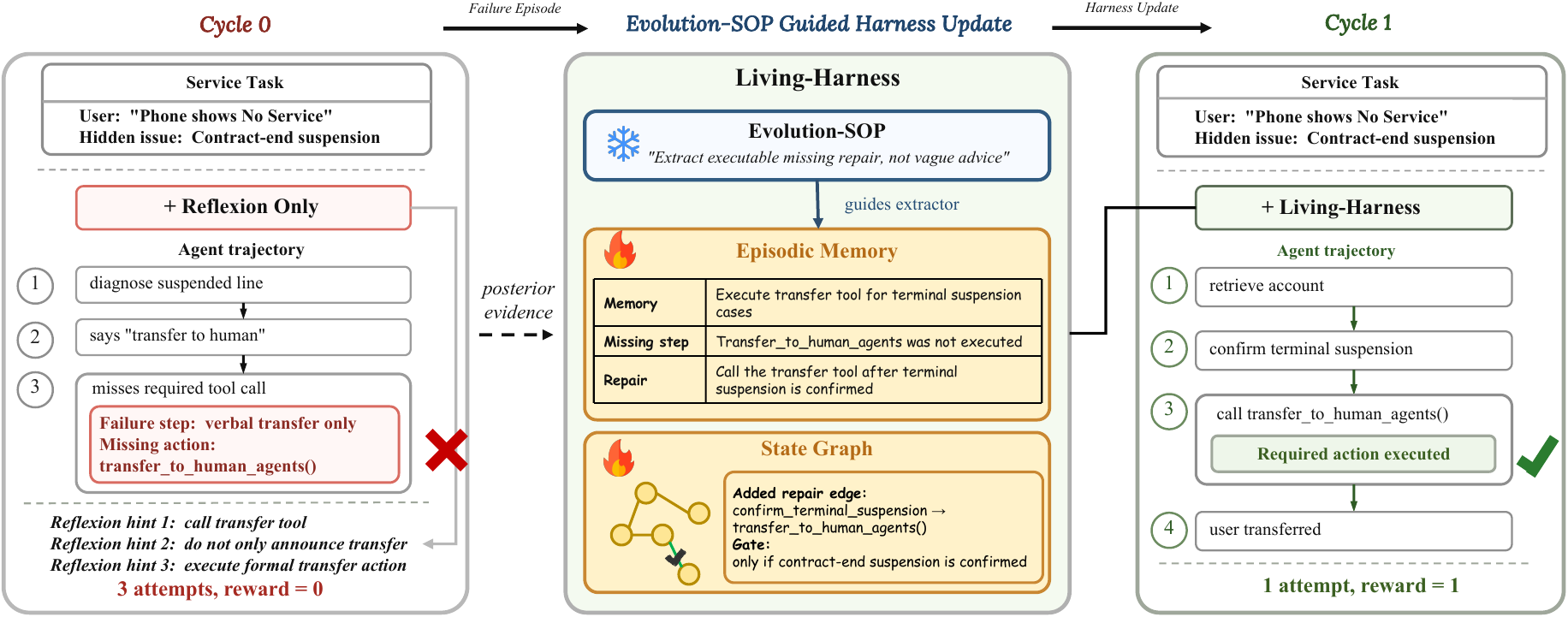}
    \caption{
    Case study of persistent procedural repair. In Cycle~0, Reflexion
    recognizes that the user should be transferred but repeatedly omits
    the required \texttt{transfer\_to\_human\_agents()} call. Guided by
    the Evolution-SOP, Living-Harness converts this evaluated failure into
    an episodic-memory entry and a state-graph repair edge. In Cycle~1,
    the repaired harness retrieves the missing procedure, executes the
    tool call, and completes the task in one attempt.
    }
    \label{fig:living-harness-repair}
\end{figure*}

\subsection{Ablation Studies}
\noindent\textbf{Ablation definitions.}
In \textit{w/o Evolution-SOP}, we keep the memory and state graph containers but replace the domain update procedure with a generic extractor, disabling domain-specific commit gates and family-scoped update rules.
In \textit{w/o Memory}, episodic-memory retrieval and memory updates are disabled.
In \textit{w/o State Graph}, graph retrieval and updates are disabled.

\noindent\textbf{Component ablation.}
Figure~\ref{fig:bar_chart} shows that all components contribute to the final performance of Living-Harness. Removing the Evolution-SOP causes the largest drop, reducing the average score from 83.09 to 73.38, suggesting that the gain is not due to attaching memory or a workflow graph alone, but from structured posterior interpretation and bounded state evolution. Removing memory or the state graph also degrades performance, with average scores of 77.34 and 79.50, respectively, indicating that reusable recovery lessons and structured workflow transitions provide complementary benefits. The full Living-Harness achieves the best result in every domain, suggesting that the complete system provides gains that are not limited to a single domain.

\subsection{In-depth Analysis}

\noindent\textbf{Harness transfer.}
Table~\ref{tab:lh_cross_model_transfer} evaluates whether a Living-Harness state evolved with GPT-5.2 on the source interaction stream can be frozen and transferred to other backbones.
During target-model evaluation, the transferred backbone only retrieves from the frozen episodic memory and state graph, with no additional global updates.
Across Gemini 3 Pro, GLM-5, Qwen3-max, and Kimi-k2, the frozen harness improves every reported domain score.
The gains are especially large in domains where the base model struggles, such as Taxi, where GLM-5, Qwen3-max, and Kimi-k2 improve from 0.00 to 43.08, 45.13, and 45.13, respectively.
Importantly, the harness also improves stronger backbones such as Gemini 3 Pro, suggesting that the learned state is not useful only for weaker models.
Instead, the memory and state graph encode reusable harness-level procedural knowledge that different actors can retrieve and exploit.

\noindent\textbf{Executable repair: a case study.}
Figure~\ref{fig:living-harness-repair} illustrates how a retrieved repair can change later agent behaviors. In the failure cycle, Reflexion repeatedly identifies the correct high-level lesson: the agent should transfer the user to a human agent, but the rollout still fails because the required tool call is never executed. Living-Harness converts this failure into an executable repair: the missing action is stored both as episodic memory and as a state graph edge linking terminal suspension detection to \texttt{transfer\_to\_human\_agents()}. The retrieved repair is not a vague advice but a state-conditioned repair that guides the required transfer action. This illustrates the key distinction from response-level reflection: recurring failures become retrievable repairs that shape future rollouts.

\section{Conclusion}

We introduced \textbf{Living-Harness}, a harness-state evolution framework that converts recurring interaction failures into bounded updates of an external executable program state. Rather than treating failures as isolated traces for response-level correction, Living-Harness uses them as evidence for revising the episodic memory and workflow structures that guide future rollouts. This shifts interactive agent adaptation from improving only the next response to maintaining a persistent, retrievable procedure around the model. Experiments on \(\tau^2\)-Bench and MultiWOZ-2.4 show improved task performance, recovery accumulation across cycles, and retrieval-only reuse across model backbones. These results suggest that reliable agents need not rely solely on stronger single-step generation or model-side updates; they can also improve by evolving the bounded harness state that organizes future interaction. More broadly, Living-Harness points toward agent systems that learn not only what to answer next, but also how future tasks should be procedurally handled while remaining grounded by priors, structure, and constraints.

\bibliography{references}

\twocolumn[
\begin{center}
{\LARGE\bfseries
Supplementary Material for\\
\textit{Living-Harness} Is an Interactive-Agent Evolver
\par}

\vspace{0.8em}

\begin{minipage}{0.92\textwidth}
\small
\textbf{Abstract.}
This supplementary document provides the formal proof, extended
reproducibility details, additional harness-state analyses, prompt
specifications, and an expanded discussion of limitations for the
main paper. The main paper is self-contained; the material below
provides supporting technical detail and extended results.
\end{minipage}
\end{center}

\vspace{1em}
]

\setcounter{secnumdepth}{2}
\renewcommand{\hgainpos}[2]{#1\,(+#2)}

\appendix

\section{Proof of Program-State Belief Tightening}
\label{app:program-state-belief}

This section gives the formal statement and proof of Proposition~1 in the main paper. The result provides an information-theoretic justification for representing the execution-guiding procedural state \(z\): adding an informative state variable refines the available information and therefore cannot increase the optimal Bayes estimation error. It does not claim that every approximate implementation of \(z\) must strictly improve performance; strict improvement requires that \(z\) contain nonzero conditional information about execution-relevant latent variables.

\paragraph{Setup.}
Let \((\Omega,\mathcal{F},\mathbb{P})\) be a probability space. Let \(\xi\in L^2(\Omega)\) denote an execution-relevant latent variable, such as the valid workflow branch, hidden failure mode, or next recoverable state. Let \(\mathcal{I}_s\subseteq\mathcal{F}\) denote the information available to an environment-only belief, and let \(\mathcal{I}_{s,z}\subseteq\mathcal{F}\) denote the refined information available after additionally modeling the procedural state \(z\). We assume the natural information-nesting condition
\begin{equation}
\mathcal{I}_s\subseteq\mathcal{I}_{s,z},
\end{equation}
which means that the augmented belief can ignore \(z\) and recover the environment-only belief.

\paragraph{Formal statement.}
For any information set \(\mathcal{I}\), define the Bayes mean-squared error bound
\begin{equation}
\mathcal{B}(\mathcal{I})
:=
\inf_{\widehat{\xi}\in L^2(\mathcal{I})}
\mathbb{E}\!\left[(\xi-\widehat{\xi})^2\right],
\end{equation}
where \(L^2(\mathcal{I})\) is the set of square-integrable \(\mathcal{I}\)-measurable estimators.

\noindent\textbf{Proposition A.1 (Formal version of Proposition~1).}
Under the setup above, the environment-only and program-augmented Bayes error bounds satisfy
\begin{equation}
\mathcal{B}(\mathcal{I}_{s,z})
=
\mathcal{B}(\mathcal{I}_s)-\Gamma_z
\le
\mathcal{B}(\mathcal{I}_s),
\end{equation}
where
\begin{equation}
\Gamma_z
=
\mathbb{E}\!\left[
\operatorname{Var}\!\left(
\mathbb{E}[\xi\mid \mathcal{I}_{s,z}]
\mid \mathcal{I}_s
\right)
\right]
\ge 0.
\end{equation}
Moreover, the inequality is strict whenever \(\Gamma_z>0\), equivalently when the refined posterior mean \(\mathbb{E}[\xi\mid\mathcal{I}_{s,z}]\) is not \(\mathcal{I}_s\)-measurable.

\paragraph{Proof.}
For any information set \(\mathcal{I}\), the Bayes estimator under squared loss is the conditional expectation \(\mu_{\mathcal{I}}=\mathbb{E}[\xi\mid\mathcal{I}]\). Hence,
\begin{equation}
\mathcal{B}(\mathcal{I})
=
\mathbb{E}\!\left[(\xi-\mu_{\mathcal{I}})^2\right]
=
\mathbb{E}\!\left[\operatorname{Var}(\xi\mid\mathcal{I})\right].
\end{equation}
Since \(\mathcal{I}_s\subseteq\mathcal{I}_{s,z}\), the conditional-variance decomposition gives
\begin{align}
\operatorname{Var}(\xi\mid\mathcal{I}_s)
&=
\mathbb{E}\!\left[
\operatorname{Var}(\xi\mid\mathcal{I}_{s,z})
\mid \mathcal{I}_s
\right] \\
&\quad+
\operatorname{Var}\!\left(
\mathbb{E}[\xi\mid\mathcal{I}_{s,z}]
\mid \mathcal{I}_s
\right).
\end{align}
Taking expectations on both sides yields
\begin{equation}
\mathcal{B}(\mathcal{I}_s)
=
\mathcal{B}(\mathcal{I}_{s,z})+\Gamma_z.
\end{equation}
Rearranging gives
\begin{equation}
\mathcal{B}(\mathcal{I}_{s,z})
=
\mathcal{B}(\mathcal{I}_s)-\Gamma_z
\le
\mathcal{B}(\mathcal{I}_s),
\end{equation}
because \(\Gamma_z\ge 0\). The inequality is strict exactly when \(\Gamma_z>0\). Since a conditional variance is zero if and only if the conditioned random variable is measurable with respect to the conditioning sigma-algebra, \(\Gamma_z>0\) holds when \(\mathbb{E}[\xi\mid\mathcal{I}_{s,z}]\) is not already determined by \(\mathcal{I}_s\). This completes the proof.

\paragraph{Approximate implementation.}
The proposition concerns the optimal Bayes estimator under a refined information set. Living-Harness realizes this refinement approximately through retrieved procedural context, posterior abstraction, and bounded state updates.

Let \(\widetilde{\mu}_{s,z}\) be an \(\mathcal{I}_{s,z}\)-measurable estimator satisfying
\begin{equation}
\mathbb{E}\!\left[
(\widetilde{\mu}_{s,z}-\mu_{\mathcal{I}_{s,z}})^2
\right]
\le
\epsilon_z.
\end{equation}
Then
\begin{equation}
\mathbb{E}\!\left[
(\xi-\widetilde{\mu}_{s,z})^2
\right]
\le
\mathcal{B}(\mathcal{I}_s)-\Gamma_z+\epsilon_z.
\end{equation}

\paragraph{Proof.}
Since \(\widetilde{\mu}_{s,z}-\mu_{\mathcal{I}_{s,z}}\) is \(\mathcal{I}_{s,z}\)-measurable and \(\mathbb{E}[\xi-\mu_{\mathcal{I}_{s,z}}\mid\mathcal{I}_{s,z}]=0\), the cross term vanishes:
\begin{align}
\mathbb{E}\!\left[
(\xi-\widetilde{\mu}_{s,z})^2
\right]
&=
\mathbb{E}\!\left[
(\xi-\mu_{\mathcal{I}_{s,z}})^2
\right] \\
&\quad+
\mathbb{E}\!\left[
(\widetilde{\mu}_{s,z}-\mu_{\mathcal{I}_{s,z}})^2
\right] \\
&\le
\mathcal{B}(\mathcal{I}_{s,z})+\epsilon_z \\
&=
\mathcal{B}(\mathcal{I}_s)-\Gamma_z+\epsilon_z.
\end{align}
Thus, the approximate augmented estimator has a lower upper bound than the environment-only Bayes bound whenever \(\epsilon_z<\Gamma_z\). More generally, relative to an approximate environment-only estimator with excess error \(\epsilon_s\), the augmented upper bound is lower whenever
\begin{equation}
\epsilon_z-\epsilon_s<\Gamma_z.
\end{equation}
This condition states that the information gain contributed by \(z\) must exceed the additional approximation error introduced by representing and using it.

\paragraph{Decision-risk version.}
The same monotonicity holds for general decision risk. Let \(\mathcal{A}\) be an action space and let \(\ell(a,\xi)\) be a loss incurred by taking action \(a\) when the execution-relevant latent variable is \(\xi\). For an information set \(\mathcal{I}\), let \(\Pi(\mathcal{I})\) denote all \(\mathcal{I}\)-measurable policies. Define the optimal Bayes risk
\begin{equation}
\mathcal{J}^{\star}(\mathcal{I})
=
\inf_{\pi\in\Pi(\mathcal{I})}
\mathbb{E}[\ell(\pi,\xi)].
\end{equation}
Since \(\mathcal{I}_s\subseteq\mathcal{I}_{s,z}\), we have \(\Pi(\mathcal{I}_s)\subseteq\Pi(\mathcal{I}_{s,z})\). Therefore,
\begin{align}
\mathcal{J}^{\star}(\mathcal{I}_{s,z})
&=
\inf_{\pi\in\Pi(\mathcal{I}_{s,z})}
\mathbb{E}[\ell(\pi,\xi)] \\
&\le
\inf_{\pi\in\Pi(\mathcal{I}_s)}
\mathbb{E}[\ell(\pi,\xi)]
=
\mathcal{J}^{\star}(\mathcal{I}_s).
\end{align}

\paragraph{Implication for Living-Harness.}
The variable \(z\) captures retrievable workflow structure, recoverable failure modes, and episodic lessons under frozen execution constraints. Proposition~A.1 shows that, when \(z\) is informative, marginalizing it away leaves avoidable uncertainty: explicitly representing \(z\) contracts the Bayes error bound by \(\Gamma_z\). The approximate result further shows that this benefit holds when the information gain from \(z\) exceeds the error introduced by extracting and using it. The practical algorithm approximates this interpretation through retrieval, extraction, and gated harness-state updates rather than exact Bayesian filtering.

\section{Reproducibility and Experimental Details}
\label{app:reproducibility-and-experimental-details}

This section provides supporting implementation details for the main paper. Section~\ref{app:online-scoring-protocol} specifies the score-before-update protocol, Section~\ref{app:retrieval-update-gates} details retrieval and commit gates, and Section~\ref{app:baseline-cost-profile} summarizes the shared baseline configuration.

\subsection{Online Scoring and Update Protocol}
\label{app:online-scoring-protocol}

For each episode \(n\), the actor first performs rollout using only the harness state available before that episode, \(S^{(n)}=(\mathcal{R}^{(n)},G^{(n)})\), together with the frozen domain context \(C_d\). The benchmark evaluator then computes the task score from the completed trajectory. Only after this score has been recorded do the posterior generator, memory extractor, and workflow extractor process the trajectory and propose updates to the global harness state.

Therefore, evidence generated from an episode is never available when computing that episode's reported score. For \(\tau^2\)-Bench, the global harness state is updated after each scored episode. For MultiWOZ-2.4, updates are synchronized every four scored episodes: all episodes in the synchronization window are first rolled out and scored, and their evidence is committed only afterwards. This prevents same-instance global-state reuse while allowing adaptation over the online interaction stream.

Task-local reflexion is handled separately from persistent harness adaptation. A local reflection buffer may guide retries within the current task instance, but it is discarded after that instance and is never directly written into the global episodic memory \(\mathcal{R}^{(n)}\) or state graph \(G^{(n)}\).

\subsection{Retrieval, Update Normalization, and Commit Gates}
\label{app:retrieval-update-gates}

At the beginning of each episode, Living-Harness retrieves context from episodic memory and the state graph. Unless otherwise specified, both retrieval modules use top-\(k=3\). Retrieval follows a same-family-first strategy: candidates from the same task family are prioritized, and cross-family candidates are used only when same-family evidence is insufficient. Candidate memories and graph fragments are ranked by semantic relevance to the current task query, task-family compatibility, and accumulated confidence or support.

Extractor outputs are normalized into schema-constrained JSON before commitment. Each update must specify its task family, trigger condition, failure pattern, proposed repair, and confidence signal. Malformed outputs receive one repair pass; an output that still violates the schema is discarded.

Global updates are committed through the Evolution-SOP-defined update pipeline. A candidate must pass the following gates:
\begin{itemize}
    \item \textbf{Schema gate}: the update must satisfy the required memory or graph schema.
    \item \textbf{Scope gate}: the update is committed to the corresponding task family unless cross-family transfer is explicitly supported.
    \item \textbf{Evidence gate}: the update must be grounded in evaluator feedback, trajectory evidence, or repeated failure patterns.
    \item \textbf{Constraint gate}: the update must not override frozen domain policies, tool preconditions, or execution constraints in \(C_d\).
    \item \textbf{Merge gate}: semantically similar updates are merged with existing entries using semantic hashes and accumulated confidence rather than inserted as conflicting duplicates.
\end{itemize}

For episodic memory, committed updates create or strengthen reusable failure--repair lessons. For the state graph, committed updates mainly merge or strengthen state-conditioned repair edges. These gates reduce the risk that noisy or misattributed evaluator signals become persistent state; they do not provide full rollback or regression testing.

\subsection{Baseline Configuration and Cost Profile}
\label{app:baseline-cost-profile}

All GPT-5.2-based interactive methods use GPT-5.2 with medium reasoning effort as the backbone. In Living-Harness, the actor and all evolution modules share this backbone. For \(\tau^2\)-Bench, the simulated user is implemented with GPT-5.1 following the benchmark setting. All interactive baselines use the same tool interface, evaluator, and task-level retry budget whenever applicable.

For Reflexion, the agent is allowed at most three trials per task, matching the task-local reflection budget used by Living-Harness. ReAct follows its standard reasoning--acting loop under the same tool and interaction budget. For memory-based baselines such as AWM and ReasoningBank, retrieved memory is inserted into the actor context under the same context-budget constraint. Unlike these baselines, Living-Harness performs post-episode abstraction and gated updates to persistent harness state only after the task has been scored.

For transfer experiments, target backbones receive retrieval-only access to the frozen source harness state, with no additional global updates during target-model evaluation.


\subsection{API-Based Computing Infrastructure}
\label{app:api-computing-infrastructure}

All model inference was conducted through remote OpenAI-compatible
API endpoints; no model was trained or hosted locally. Consequently,
the provider-side GPU/CPU models, accelerator memory, operating system,
and serving software were not exposed to us. The local runtime was
used only for agent orchestration, tool execution, retrieval,
harness-state persistence, and metric aggregation. We therefore report
the model identifiers and inference configurations used in the
experiments, while treating the provider-side computing infrastructure
as unavailable.

\section{Additional Results}
\label{app:additional-results}

\paragraph{Harness-state scaling.}
Tables~\ref{tab:tau2-state-scaling} and~\ref{tab:multiwoz-state-scaling} report how the persistent harness state grows over evolution cycles. Cycle~0 denotes the empty pre-evolution state. For each domain, \(|\mathcal{R}^{(n)}|\) is the number of episodic-memory entries after cycle \(n\), while \(|V_G^{(n)}|\) and \(|E_G^{(n)}|\) denote the numbers of nodes and edges in the state graph. Parenthesized values report absolute increases over the preceding cycle.

\begin{table*}[t]
\centering
\small
\setlength{\tabcolsep}{6pt}
\begin{tabular*}{0.94\textwidth}{@{\extracolsep{\fill}}lcccc@{}}
\toprule
Metric & Cycle 0 & Cycle 1 & Cycle 2 & Cycle 3 \\
\midrule
\multicolumn{5}{l}{\textit{Airline}} \\
\midrule
\(|\mathcal{R}^{(n)}|\) & 0 & \hgainpos{26}{26} & \hgainpos{52}{26} & \hgainpos{71}{19} \\
\(|V_G^{(n)}|\) & 0 & \hgainpos{197}{197} & \hgainpos{336}{139} & \hgainpos{471}{135} \\
\(|E_G^{(n)}|\) & 0 & \hgainpos{200}{200} & \hgainpos{400}{200} & \hgainpos{600}{200} \\
\midrule
\multicolumn{5}{l}{\textit{Retail}} \\
\midrule
\(|\mathcal{R}^{(n)}|\) & 0 & \hgainpos{67}{67} & \hgainpos{131}{64} & \hgainpos{198}{67} \\
\(|V_G^{(n)}|\) & 0 & \hgainpos{304}{304} & \hgainpos{521}{217} & \hgainpos{714}{193} \\
\(|E_G^{(n)}|\) & 0 & \hgainpos{456}{456} & \hgainpos{912}{456} & \hgainpos{1368}{456} \\
\midrule
\multicolumn{5}{l}{\textit{Telecom}} \\
\midrule
\(|\mathcal{R}^{(n)}|\) & 0 & \hgainpos{54}{54} & \hgainpos{107}{53} & \hgainpos{166}{59} \\
\(|V_G^{(n)}|\) & 0 & \hgainpos{308}{308} & \hgainpos{547}{239} & \hgainpos{715}{168} \\
\(|E_G^{(n)}|\) & 0 & \hgainpos{365}{365} & \hgainpos{719}{354} & \hgainpos{1030}{311} \\
\bottomrule
\end{tabular*}
\caption{Harness-state scaling on \(\tau^2\)-Bench. Cycle~0 is the empty pre-evolution state; Cycles~1--3 report accumulated episodic memory and state-graph size after each evolution cycle. Parenthesized values are absolute increases from the preceding cycle.}
\label{tab:tau2-state-scaling}
\end{table*}

\begin{table*}[t]
\centering
\small
\setlength{\tabcolsep}{6pt}
\begin{tabular*}{0.94\textwidth}{@{\extracolsep{\fill}}lcccc@{}}
\toprule
Metric & Cycle 0 & Cycle 1 & Cycle 2 & Cycle 3 \\
\midrule
\multicolumn{5}{l}{\textit{Restaurant}} \\
\midrule
\(|\mathcal{R}^{(n)}|\) & 0 & \hgainpos{333}{333} & \hgainpos{647}{314} & \hgainpos{951}{304} \\
\(|V_G^{(n)}|\) & 0 & \hgainpos{326}{326} & \hgainpos{535}{209} & \hgainpos{704}{169} \\
\(|E_G^{(n)}|\) & 0 & \hgainpos{590}{590} & \hgainpos{1013}{423} & \hgainpos{1363}{350} \\
\midrule
\multicolumn{5}{l}{\textit{Hotel}} \\
\midrule
\(|\mathcal{R}^{(n)}|\) & 0 & \hgainpos{239}{239} & \hgainpos{479}{240} & \hgainpos{729}{250} \\
\(|V_G^{(n)}|\) & 0 & \hgainpos{357}{357} & \hgainpos{585}{228} & \hgainpos{799}{214} \\
\(|E_G^{(n)}|\) & 0 & \hgainpos{585}{585} & \hgainpos{1040}{455} & \hgainpos{1454}{414} \\
\midrule
\multicolumn{5}{l}{\textit{Train}} \\
\midrule
\(|\mathcal{R}^{(n)}|\) & 0 & \hgainpos{214}{214} & \hgainpos{435}{221} & \hgainpos{669}{234} \\
\(|V_G^{(n)}|\) & 0 & \hgainpos{291}{291} & \hgainpos{438}{147} & \hgainpos{558}{120} \\
\(|E_G^{(n)}|\) & 0 & \hgainpos{621}{621} & \hgainpos{1015}{394} & \hgainpos{1373}{358} \\
\midrule
\multicolumn{5}{l}{\textit{Attraction}} \\
\midrule
\(|\mathcal{R}^{(n)}|\) & 0 & \hgainpos{47}{47} & \hgainpos{102}{55} & \hgainpos{167}{65} \\
\(|V_G^{(n)}|\) & 0 & \hgainpos{361}{361} & \hgainpos{573}{212} & \hgainpos{759}{186} \\
\(|E_G^{(n)}|\) & 0 & \hgainpos{633}{633} & \hgainpos{1084}{451} & \hgainpos{1475}{391} \\
\midrule
\multicolumn{5}{l}{\textit{Taxi}} \\
\midrule
\(|\mathcal{R}^{(n)}|\) & 0 & \hgainpos{68}{68} & \hgainpos{146}{78} & \hgainpos{237}{91} \\
\(|V_G^{(n)}|\) & 0 & \hgainpos{94}{94} & \hgainpos{140}{46} & \hgainpos{192}{52} \\
\(|E_G^{(n)}|\) & 0 & \hgainpos{186}{186} & \hgainpos{291}{105} & \hgainpos{394}{103} \\
\bottomrule
\end{tabular*}
\caption{Harness-state scaling on MultiWOZ-2.4. Cycle~0 is the empty pre-evolution state; Cycles~1--3 report accumulated state size after each evolution cycle. Parenthesized values are absolute increases from the preceding cycle.}
\label{tab:multiwoz-state-scaling}
\end{table*}

\paragraph{Comparison with interactive and self-evolving baselines.}
Table~\ref{tab:method-comparison} contrasts Living-Harness with representative baselines along the dimensions most relevant to harness evolution: the object being updated, whether persistent state is retained, whether the external procedure is revised, and whether updates are bounded by explicit commit gates.

\begin{table*}[t]
\centering
\small
\setlength{\tabcolsep}{5pt}
\begin{tabular}{@{}p{0.14\textwidth}p{0.28\textwidth}p{0.14\textwidth}p{0.14\textwidth}p{0.16\textwidth}@{}}
\toprule
\textbf{Method} & \textbf{Update object} & \textbf{Persistent state} & \textbf{Procedure revised} & \textbf{Bounded gates} \\
\midrule
ReAct & Trajectories & No & No & No \\
Reflexion & Verbal memories & Yes & No & No \\
AWM & Workflows & Yes & Partial & No \\
ReasoningBank & Reasoning strategies & Yes & No & No \\
EvoTest & Agent configurations & Yes & Yes & No \\
Meta-Harness & Harness code & Yes & Yes & No \\
\textbf{Living-Harness} & Episodic memories and state graph & Yes & Yes & Yes, through the Evolution-SOP \\
\bottomrule
\end{tabular}
\caption{Comparison with interactive and self-evolving baselines. ``Partial'' denotes a method that revises reusable workflow content without maintaining the jointly gated memory-and-graph state used by Living-Harness.}
\label{tab:method-comparison}
\end{table*}

\section{Limitations and Broader Impacts}
\label{app:limitations-and-broader-impacts}

\paragraph{Evaluation scope.}
Living-Harness is evaluated in controlled interactive-agent benchmarks, and the empirical conclusions should be interpreted within this scope. Although \(\tau^2\)-Bench and MultiWOZ-2.4 cover multiple transactional and dialogue settings, they remain simulator-based environments with structured task definitions, finite tool sets, and evaluator-defined success criteria. The score-before-update protocol prevents same-instance global-state reuse, but does not establish robustness to arbitrary stream orders, reduced recurrence, unseen policy changes, or open-ended real-world deployments. Future work should evaluate shuffled streams, held-out task families, policy perturbations, and less structured user goals.

\paragraph{State reliability.}
The schema, scope, evidence, constraint, and merge gates reduce malformed, unsupported, or conflicting updates, but they do not guarantee monotonic improvement. The current system does not implement full rollback, systematic stale-entry removal, or regression testing over previously solved tasks. Incorrect evaluator feedback or overly narrow repairs may therefore persist and affect later retrieval.

\paragraph{Portability and manual design.}
The posterior--extract--commit pipeline and update schemas are shared across domains, whereas each Evolution-SOP instantiates run-frozen domain monitoring and task-family scope rules. These domain rules are currently specified manually. The present experiments do not establish zero-shot transfer of an Evolution-SOP to unseen domains or the performance of a fully generic update policy.

\paragraph{Cost, privacy, and deployment considerations.}
Living-Harness introduces post-episode computation for posterior generation, memory extraction, workflow extraction, schema normalization, retrieval, and state commitment. These costs are separated from scored rollout computation but remain relevant in deployment. Persistent interaction histories may also contain privacy-sensitive information; practical deployments should apply data minimization, access control, retention policies, and redaction appropriate to the application. Additional safeguards may be needed for noisy tools, ambiguous user goals, policy changes, and incorrect evaluator feedback.

\section{Prompt Specifications}
\label{app:prompts}

\begin{figure*}[t]
\centering
\includegraphics[width=0.94\textwidth]{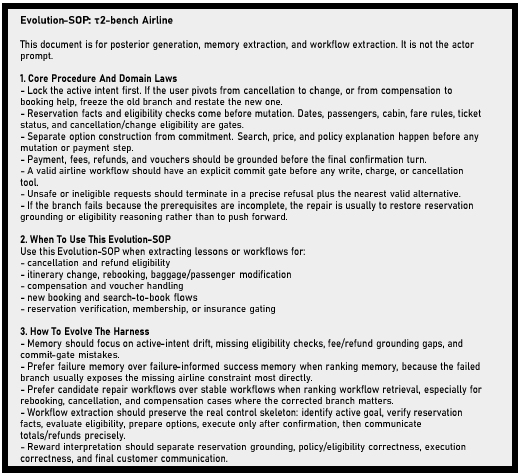}
\caption{\(\tau^2\)-Bench Airline Evolution-SOP: The run-frozen Evolution-SOP for the airline domain specifies monitoring rules for interpreting evaluated trajectories, diagnosing recurring failures, and constraining memory and state-graph updates.}
\label{fig:prompt_tau2_airline_meta_sop}
\end{figure*}

\begin{figure*}[t]
\centering
\includegraphics[width=0.94\textwidth]{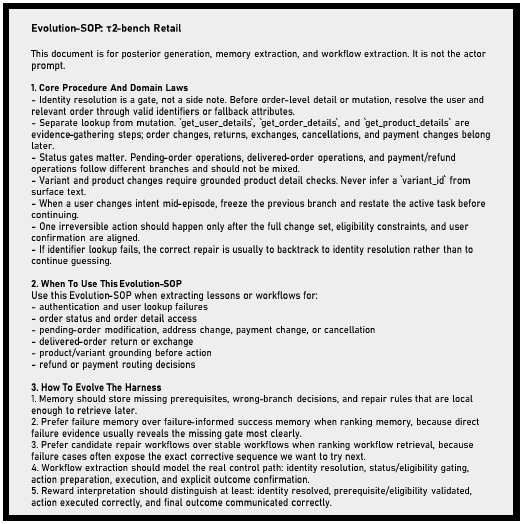}
\caption{\(\tau^2\)-Bench Retail Evolution-SOP: The run-frozen Evolution-SOP for the retail domain defines how post-episode evidence is interpreted and which reusable repairs are eligible for bounded harness-state updates.}
\label{fig:prompt_tau2_retail_meta_sop}
\end{figure*}

\begin{figure*}[t]
\centering
\includegraphics[width=0.94\textwidth]{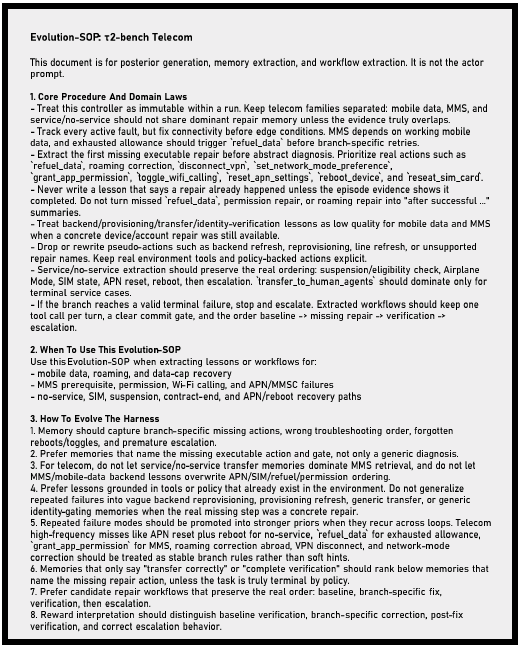}
\caption{\(\tau^2\)-Bench Telecom Evolution-SOP: The run-frozen Evolution-SOP for the telecom domain guides failure diagnosis, family isolation, and bounded updates to episodic memory and the state graph.}
\label{fig:prompt_tau2_telecom_meta_sop}
\end{figure*}

\begin{figure*}[t]
\centering
\includegraphics[width=0.94\textwidth]{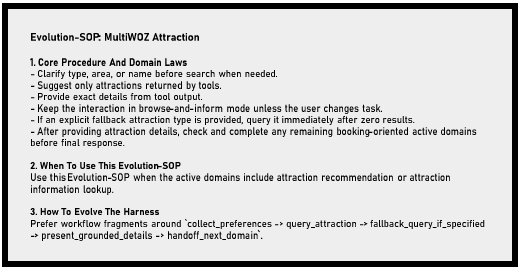}
\caption{MultiWOZ-2.4 Attraction Evolution-SOP: The domain-specific Evolution-SOP for attraction-related dialogue tasks specifies how evaluated interaction evidence is converted into reusable memory and workflow repairs.}
\label{fig:prompt_multiwoz_attraction_meta_sop}
\end{figure*}

\begin{figure*}[t]
\centering
\includegraphics[width=0.94\textwidth]{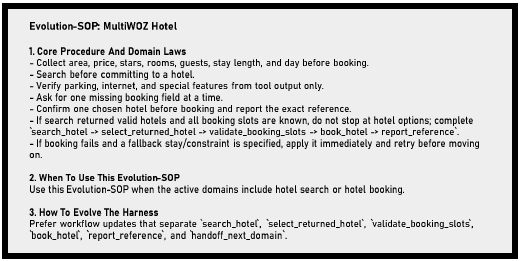}
\caption{MultiWOZ-2.4 Hotel Evolution-SOP: The domain-specific Evolution-SOP for hotel tasks defines monitoring rules for posterior interpretation and bounded harness evolution.}
\label{fig:prompt_multiwoz_hotel_meta_sop}
\end{figure*}

\begin{figure*}[t]
\centering
\includegraphics[width=0.94\textwidth]{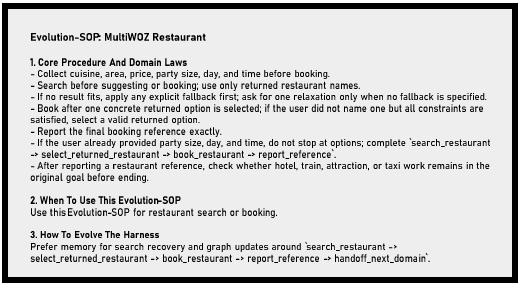}
\caption{MultiWOZ-2.4 Restaurant Evolution-SOP: The domain-specific Evolution-SOP for restaurant tasks guides evidence extraction, failure abstraction, and constrained updates to the procedural harness state.}
\label{fig:prompt_multiwoz_restaurant_meta_sop}
\end{figure*}

\begin{figure*}[t]
\centering
\includegraphics[width=0.94\textwidth]{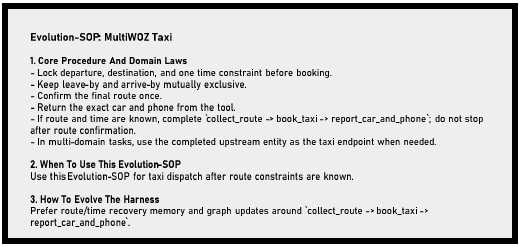}
\caption{MultiWOZ-2.4 Taxi Evolution-SOP: The domain-specific Evolution-SOP for taxi tasks specifies how post-episode failures and successful repairs are interpreted for memory and state-graph updates.}
\label{fig:prompt_multiwoz_taxi_meta_sop}
\end{figure*}

\begin{figure*}[t]
\centering
\includegraphics[width=0.94\textwidth]{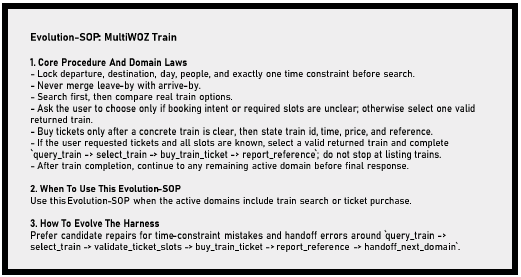}
\caption{MultiWOZ-2.4 Train Evolution-SOP: The domain-specific Evolution-SOP for train tasks defines monitoring and bounded-update rules for long-horizon dialogue interactions.}
\label{fig:prompt_multiwoz_train_meta_sop}
\end{figure*}

\begin{figure*}[t]
\centering
\includegraphics[width=0.94\textwidth]{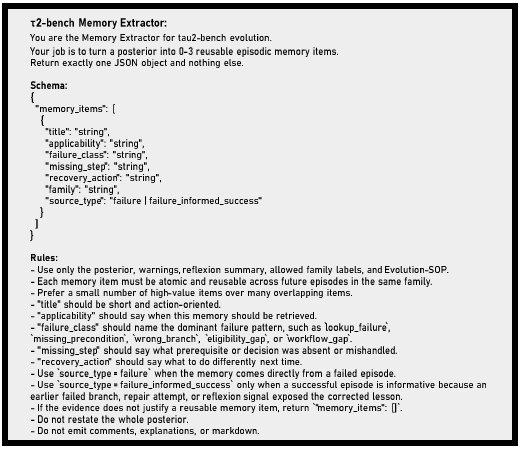}
\caption{\(\tau^2\)-Bench Memory Extractor Prompt. The memory-extractor prompt converts posterior evidence into structured episodic-memory items containing trigger conditions, failure patterns, and recovery actions.}
\label{fig:prompt_tau2_memory_extractor}
\end{figure*}

\begin{figure*}[t]
\centering
\includegraphics[width=0.94\textwidth]{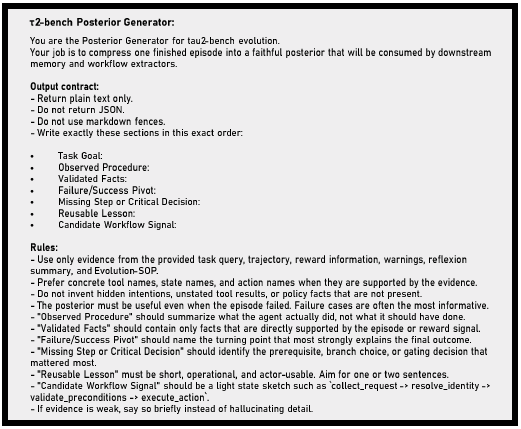}
\caption{\(\tau^2\)-Bench Posterior Generator Prompt. The posterior-generator prompt compresses evaluated trajectories and evaluator signals into episode-level evidence for subsequent memory and workflow extraction.}
\label{fig:prompt_tau2_posterior_generator}
\end{figure*}

\begin{figure*}[t]
\centering
\includegraphics[width=0.94\textwidth]{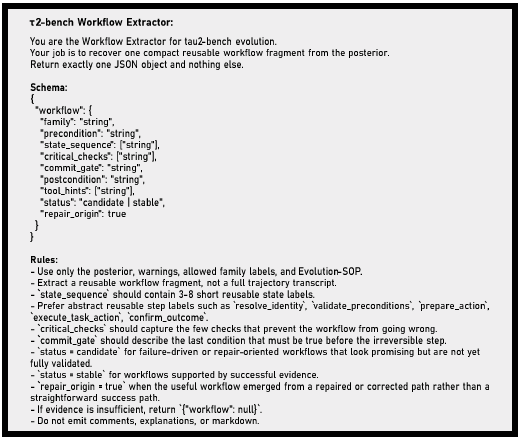}
\caption{\(\tau^2\)-Bench Workflow Extractor Prompt. The workflow-extractor prompt maps posterior evidence into state-graph updates such as state nodes, repair edges, and transition rules.}
\label{fig:prompt_tau2_workflow_extractor}
\end{figure*}

\begin{figure*}[t]
\centering
\includegraphics[width=0.94\textwidth]{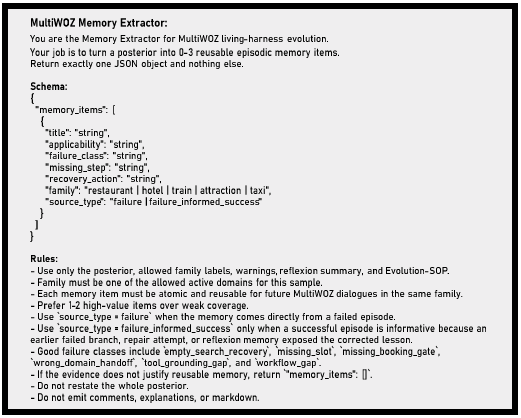}
\caption{MultiWOZ-2.4 Memory Extractor Prompt. The memory-extractor prompt stores reusable dialogue-level failure patterns and recovery actions as structured episodic memory.}
\label{fig:prompt_multiwoz_memory_extractor}
\end{figure*}

\begin{figure*}[t]
\centering
\includegraphics[width=0.94\textwidth]{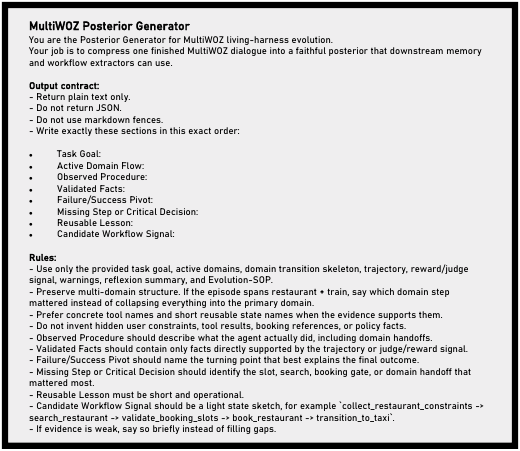}
\caption{MultiWOZ-2.4 Posterior Generator Prompt. The posterior-generator prompt abstracts evaluated dialogue trajectories into evidence for bounded harness evolution.}
\label{fig:prompt_multiwoz_posterior_generator}
\end{figure*}

\begin{figure*}[t]
\centering
\includegraphics[width=0.94\textwidth]{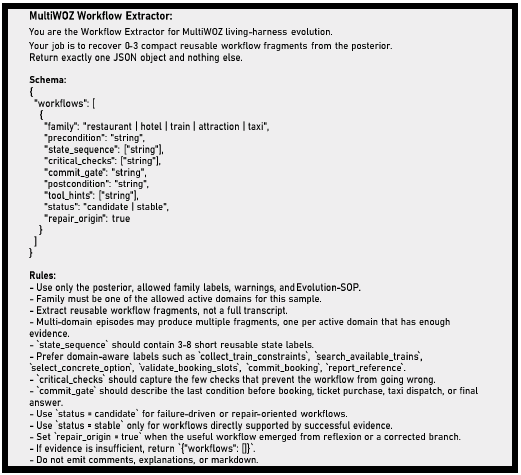}
\caption{MultiWOZ-2.4 Workflow Extractor Prompt. The workflow-extractor prompt converts dialogue-level evidence into state-graph refinements for future retrieval and execution.}
\label{fig:prompt_multiwoz_workflow_extractor}
\end{figure*}

This section provides the prompt specifications used by Living-Harness. Figures~\ref{fig:prompt_tau2_airline_meta_sop}--\ref{fig:prompt_multiwoz_train_meta_sop} show the run-frozen Evolution-SOPs for the three \(\tau^2\)-Bench domains and five MultiWOZ-2.4 domains. Figures~\ref{fig:prompt_tau2_memory_extractor}--\ref{fig:prompt_multiwoz_workflow_extractor} show the posterior-generation, memory-extraction, and workflow-extraction prompts for the two benchmark families. These prompts instantiate the update mechanism described in the main paper.

\end{document}